\documentclass[%
  aps,
  rmp,
 ]{revtex4-1}

\usepackage[utf8]{inputenc}
\usepackage[T1]{fontenc}
\usepackage[english]{babel}
\usepackage{amsmath,amssymb,amsfonts,amsthm}
\usepackage{bm,dsfont,mathtools}
\usepackage{graphicx}
\usepackage{hyperref}
\usepackage{braket}

\setcitestyle{numbers,square,sort&compress}

\begin{document}

\newcommand{\hilb}[1]{\mathcal{H}^{#1}} 
\newcommand{\trc}[0]{\text{tr}} 
\newcommand{\rk}[0]{\text{rk}} 
\newcommand{\dimn}[0]{\text{dim}} 
\newcommand{\qeds}{\hfill\ensuremath{\blacksquare}} 
\newcommand{\eqU}[0]{\cong} 
\newcommand{\lth}[0]{\ell} 

\renewcommand{\figurename}{Fig.}
\renewcommand{\tablename}{Tab.}
\DeclarePairedDelimiter{\ceil}{\lceil}{\rceil}

\graphicspath{{figs/}}

\newtheorem{defi}{Definition}
\newtheorem{prop}{Proposition}
\newtheorem{teor}{Theorem}

\title[]{Classical extension of quantum-correlated separable states}%

\author{G. Bellomo}
    \email{gbellomo@fisica.unlp.edu.ar}
    \affiliation{Instituto de Física La Plata (IFLP-CONICET),
        and Departamento de Física, Facultad de Ciencias Exactas,
        Universidad Nacional de La Plata, 115 and 49, C.C. 67,
        1900 La Plata, Argentina}

\author{A.R. Plastino}
    \affiliation{CeBio and Secretaría de Investigaciones,
        Universidad Nacional del Noroeste de la
        Prov. de Buenos Aires (UNNOBA-CONICET),
        R. Saenz Peña 456, Junín, Argentina}

\author{A. Plastino}
    \affiliation{Instituto de Física La Plata (IFLP-CONICET),
        and Departamento de Física, Facultad de Ciencias Exactas,
        Universidad Nacional de La Plata, 115 and 49, C.C. 67,
        1900 La Plata, Argentina}

\date{\today}%

\begin{abstract}
Li and Luo [\textit{Phys. Rev. A} \textbf{78} (2008),  024303]
discovered a remarkable relation between discord and
entanglement. It establishes that all separable states can be obtained
via reduction of a classicaly-correlated state `living' in a space of
larger dimension. Starting from this result, we discuss here an
\textit{optimal classical extension} of separable states and
explore this notion for low-dimensional systems. We find that
the larger the dimension of the classical extension,
 the larger the discord in the original separable state.
 Further, we analyze separable states of maximum discord in
${\mathbb{C}^2\otimes\mathbb{C}^2}$ and their associated classical
extensions showing that, from the reduction of a classical state
in ${(\mathbb{C}^2\otimes\mathbb{C}^3)\otimes\mathbb{C}^2}$, one
can obtain a separable state of maximum discord in
${\mathbb{C}^2\otimes\mathbb{C}^2}$.
\end{abstract}

\maketitle

\section{Introduction}
Entanglement and discord are known to be quantum resources for
implementing information-computation protocols (ICP) with a higher
efficiency than that attainable via classical resources (for a
complete review see~\cite{Horo09,Modi12} and references
therein). The entanglement usefulness for such protocols has been
extensively documented. As for discord's, one can cite, for
instance,~\cite{DSC08,LBAW08,MaDa11,CABM11,GCAS12,Daki12,CVDC03,Pira13,Madh13,Horo14},
although some controversy arises regarding its
\textit{ICP-necessity}~\cite{Gess12,Gior13,StKB13}. It is clear,
however, that entanglement and discord capture different features
of the quantum world. Discord captures the fact that all classical
states must be information-wise accessible to local observers.
Thus, it is accepted that the dichotomy cassical/non-classical can
be treated in similar fashion as that regarding
discord/no-discord. For a bipartite system one associates a
Hilbert space ${\hilb{AB}=\hilb{A}\otimes\hilb{B}}$. A system's
state is represented by a positive semi-definite, hermitic
operator of trace unity acting on $\hilb{AB}$. If ${\{\Pi_i^A\}}$
and ${\{\Pi_i^B\}}$ are complete projective measurements over
$\hilb{A}$ and $\hilb{B}$, respectively, then~\cite{Luo08,Modi12}
    \begin{itemize}
    \item If ${\sigma^{AB}=\sum_i{p_i \Pi^A_i \otimes \rho^B_i}}$,
    the state is \textit{classical-quantum} (CQ): there exists a basis
    in $A$ for which the locally accessible information is maximal
    and, for an external observer, such information can be obtained
    without perturbing the composite system;
    \item If ${\sigma^{AB}=\sum_{i,j}{p_{ij} \Pi_i^A \otimes \Pi^B_j}}$,
    the state is \textit{classical-classical} (CC): the locally accessible
    information is maximal for $A$ and $B$, can be obtained without
    perturbing the composite system.
    \end{itemize}
In analogous fashion, one defines \textit{quantum-classical} (QC) states
via interchange of $A$ and $B$. We will generically speak
of classical states when referring to any of these three sub-types.
Moreover, we will speak of the set $\mathcal{CC}$ of classical-classical states,
the set $\mathcal{CQ}$ ($\mathcal{QC}$) of {classical-quantum}
({quantum-classical}) states, and the set $\mathcal{S}$ of separable states.

From the above definitions one easily ascertains that, even if the
sets $\mathcal{CC}$ and $\mathcal{CQ}$ are included within the convex
$\mathcal{S}$, neither $\mathcal{CC}$ nor $\mathcal{CQ}$ (or $\mathcal{QC}$)
constitute a convex set by themselves. Precisely, this lack of convexity
implies the existence of classical states that, via mixing amongst themselves,
may give rise to non-classical states, endowed with discord~\cite{LiLu08}. This
fact underlies the link between separability and classicality observed
by Li and Luo in~\cite{LiLu08}: a bipartite state is separable iff it can
be obtained as the reduction of a CS of larger dimension, respecting the
original bipartition. This assertion is the source of the present investigation.

Herefrom, we speak of CS when referring to CQ states. Thus, given
a bipartite state  $\rho^{ab}$, with ${\rho^a:=\trc_b[\rho^{ab}]}$
and ${\rho^b:=\trc_a[\rho^{ab}]}$, we compute the discord~\cite{OlZu01}
    \begin{equation} \label{eq:def_discord}
    \delta_a(\rho^{ab}) := I(\rho^{ab}) - C(\rho^{ab}) \,,
    \end{equation}
where
    \begin{equation} \label{eq:def_infomutua}
    I(\rho^{ab}) := S(\rho^a) + S(\rho^b) - S(\rho^{ab}),
    \end{equation}
is the quantum mutual information of the bipartite state and
    \begin{equation} \label{eq:def_infoclas}
    C(\rho^{ab}) := S(\rho^b) - \min_{ \{M^a_i\}}{ S(\rho^b|\rho^a) }
    \end{equation}
is the corresponding classical information for that state. $S(\cdot)$
is von Neumann's entropy (logarithms of basis $2$) and $\{M^a_i\}$ a
set of positive operators representing a generalized measurement over
$\hilb{a}$. ${S(\rho^b|\rho^a):=S(\rho^{ab})-S(\rho^a)}$ is the
conditional quantum entropy.

In Section~\ref{sec:sep_vs_clas} we shall introduce and adapt the Li-Luo's
relation between separability and classicality to our present
needs. We introduce in Sect.~\ref{sec:ext_opt} the notion of
\textit{optimum classical extension} for separable states as that
of smallest dimensionality. In Sect.~\ref{sec:2qubits} we will
show that, for low-dimensionality separable states
(${\mathbb{C}^2\otimes\mathbb{C}^2}$), it is possible to find
classical extensions that improve on the ones in~\cite{LiLu08}. Some explicit examples will be given. In Sect.~\ref{sec:mdss}
we discuss the existence of separable states with
maximum discord in arbitrary dimensions and consider their
relation with the notions of: (i) mutually unbiased basis and (ii)
generalized measurements that are both symmetric and
informationally complete. Some conclusions are drawn in Sect.~\ref{sec:conc}.

\section{Separability vs. Classicality: Li-Luo's relation \label{sec:sep_vs_clas}}
Monogamy is a fundamental feature of entanglement. Given a
multipartite system, if two of its parties are maximally
entangled, then they can not be entangled with a third party.
Given a composite state  $\sigma^{AB}$, with $\{A_i\}$ and
$\{B_j\}$ parts of $A$ and $B$, respectively, a monogamous
entanglement measure $E$ is such that~\cite{Terh04,Koas04}
    \begin{equation} \label{eq:ent_mono}
    E(A:B) \geq \sum_{i,j} {E(A_i:B_j)} \,,
    \end{equation}
where $E(x:y)$ yields the entanglement between  $x$ and  $y$,
${E\geq0}$. It follows from~\eqref{eq:ent_mono} that, given
$\sigma^{AB}$ not entangled, none of its reductions will exhibit
entanglement,  i.e.,
    \begin{equation} \label{eq:ent_mono2}
    E(A:B) = 0 \;\Rightarrow\; E(A_i:B_j) = 0 \;\; \forall\, i,j \,.
    \end{equation}
Reciprocally, an entangled state $\rho^{A_iB_j}$ can not be
extended to a non-entangled one $\sigma^{AB}$. The vocable
\textit{extension} will be the subject of the precise
definition~\ref{def:ext} below.
    \begin{figure}
    \centering
    \includegraphics[width=7cm]{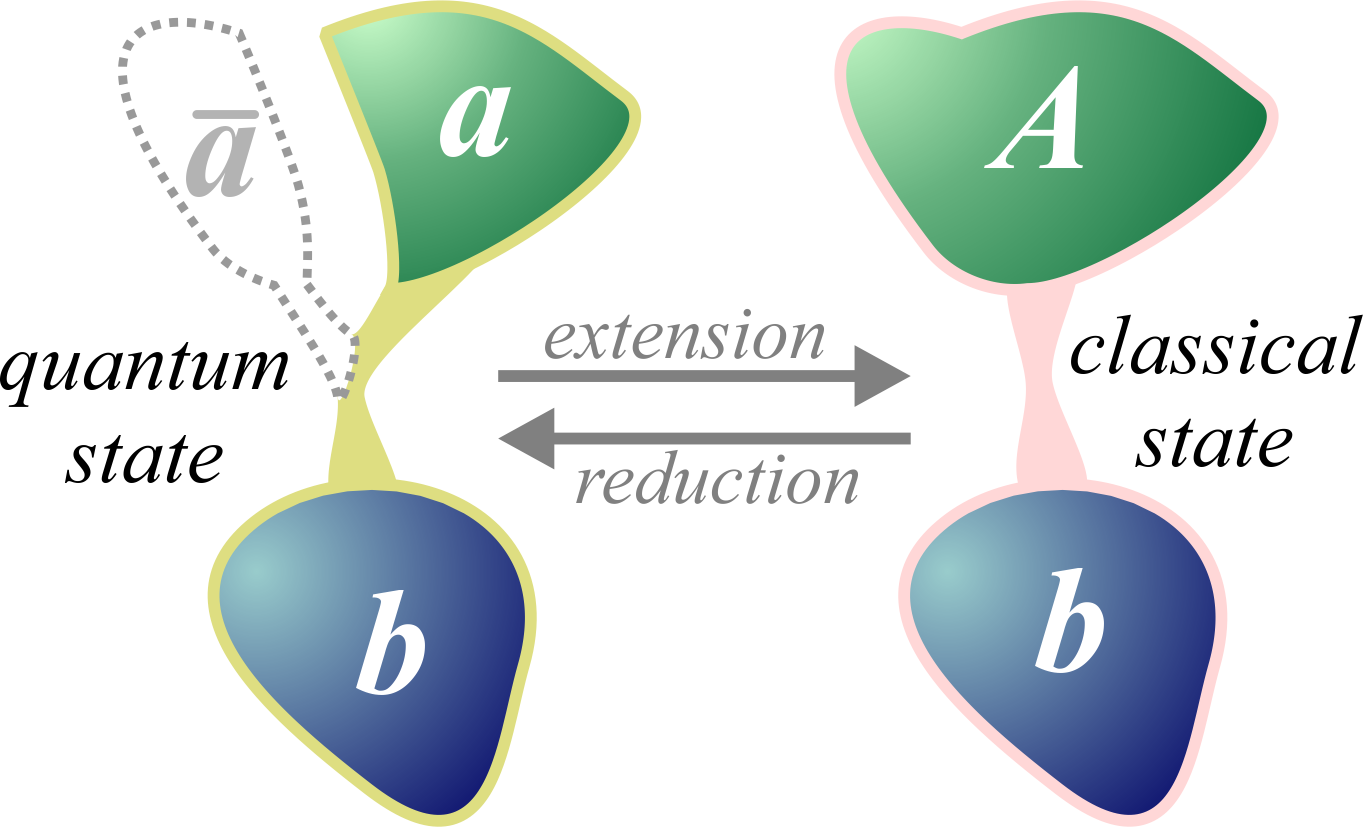}
    \caption{A separable state can always be obtained as the reduction
    of a classically correlated state embedded in a space of larger
    dimension. Li-Luo's extension algorithm provides the manner in which
    to determine the classical extension of any given separable state.}
    \label{fig:luo_scheme}
    \end{figure}

In general, discord does not obey inequalities of the
type~\eqref{eq:ent_mono}~\cite{SAPB12,PPDS12,Gior11,BMPP14}. Li
and Luo showed that any bipartite separable state \textit{can} be
extended to a CC state in a space of larger dimension~\cite{LiLu08}
(Fig.~\ref{fig:luo_scheme}). They studied
the {`separable$\rightarrow$classical-classical'} extension. We,
instead, are here interested in the
{separable$\rightarrow$classical-quantum} extension.
The following theorem explains just how to find the desired
extension~\cite{LiLu08}:
    \begin{teor} \label{teo:Luo}
    A bipartite state $\rho^{ab}$ is separable in
    ${\hilb{ab}=\hilb{a}\otimes\hilb{b}}$
    iff there exists a CQ state
    $\sigma^{Ab}$ in ${\hilb{Ab}=\hilb{A}\otimes\hilb{b}}$,
    with ${\hilb{A}=\hilb{a}\otimes\hilb{\bar{a}}}$ such that
        \begin{equation} \label{eq:luo}
        \rho^{ab}=\trc_{\bar{a}}[\sigma^{Ab}] \,.
        \end{equation}
Here, $\hilb{\bar{a}}$ is an auxiliary Hilbert space for party
$a$, while $\trc_{\bar{a}}$ is the partial trace over
$\hilb{\bar{a}}$.
    \end{teor}

    \begin{proof}
The demonstration is adapted from~\cite{LiLu08}. We start with an
arbitrary separable state
    \begin{equation} \label{eq:luo1}
    \rho^{ab}=\sum_{k=1}^{K}{p_k\rho^a_k\otimes\rho^b_k} \,.
    \end{equation}
Each $\rho^a_k$ can be expanded in its eigen-basis
$\{\ket{\alpha_{ku}}\}$ so that~\eqref{eq:luo1} can be cast as
    \begin{equation} \label{eq:luo2}
    \rho^{ab}=\sum_{k}\sum_{u}{p_k a_{ku} P^a_{ku}\otimes \rho^b_k} \,,
    \end{equation}
where we define ${P^a_{ku}:=\ket{\alpha_{ku}}\bra{\alpha_{ku}}}$.
Our extension demands consideration of an auxiliary system
$\bar{a}$, defined in ${\hilb{\bar{a}}=\mathbb{C}^K}$, such that
$\{\ket{k}\}$, with ${k=1,2...K}$ an orthonormal basis of $\mathbb{C}^K$.
Then,
    \begin{equation} \label{eq:luo3}
    \{\Pi_{ku}^A:=\ket{k}\bra{k}\otimes P^a_{ku}\} \,,
    \end{equation}
is an orthogonal set of ${\hilb{A}:=\hilb{\bar{a}}\otimes\hilb{a}}$.
Extension to a complete projective measurement in the extended space
is feasible. Define the extended state
(in ${(\mathbb{C}^K\otimes\hilb{a})\otimes\hilb{b}}$)
    \begin{equation} \label{eq:luo4}
    \sigma^{Ab}:=\sum_{k,u}{ p_k a_{ku} \Pi^A_{ku} \otimes \rho^b_{k} } \,,
    \end{equation}
 a CQ state with respect to the partition $(A,b)$.
 From its reduction one gets the separable state $\rho^{ab}$. Accordingly,
    \begin{equation} \label{eq:luo5}
    \trc_{\bar{a}}{[\sigma^{Ab}]}
        = \sum_{k,u}{ p_k a_{ku} P^a_{ku}\otimes \rho^b_{kv} }
        =\rho^{ab} \,,
    \end{equation}
as we wished to show.
    \end{proof}
This classical-extension construction-process will be called,
either \textit{Li-Luo's extension} or \textit{Li-Luo's algorithm} (LLA).
Some observations are in order.
    \begin{itemize}
        \item Our extension depends on the separable decomposition
    of the original state (see~\eqref{eq:luo1}). The party one wishes
    to make classical is extended using an ancilla in $\mathbb{C}^K$,
    with $K$ the number of terms in the decomposition.
    Luo \textit{et al.} want instead a CC state which
    needs two ancillae (one per party) in $\mathbb{C}^K$.
        \item Extending party $b$ does not change its classical nature
    when `observed' from $a$ (with a local measurement on $a$). The
    $b$-extension does not modify the classical-quantum character of the
    bipartite system. Conversely, assume the existence of a classical
    extension $\omega^{aB}$ in ${\hilb{aB}=\hilb{a}\otimes\hilb{B}}$,
    with ${\hilb{B}:=\hilb{b}\otimes\hilb{\bar{b}}}$, compatible with
    a given separable state $\rho^{ab}$. In such a case, we can write
    \begin{equation} \label{eq:item2}
        \omega^{aB} = \sum_{m,n}{ \gamma_{mn} \Pi^a_m \otimes \omega^B_n } \,,
    \end{equation}
    and tracing over the ancilla we obtain the classical state
    ${\rho^{ab}=\sum_{m,n}{ \gamma_{mn} \Pi^a_m \otimes \omega^b_n}}$,
    with ${\omega^b:=\trc_{\bar{b}}{[\omega^{B}]}}$. Thus, $\rho^{ab}$
    can not be an arbitrary separable state (it is classical).
        \item The LLA does not entangle the ancilla
    with the original system, that is, $a$ with $\bar{a}$. Actually, from~\eqref{eq:luo3}
    and~\eqref{eq:luo4} it follows that
    \begin{equation} \label{eq:item3}
    \sigma^{a\bar{a}} = \trc_{b}{[\sigma^{Ab}]}
                    = \sum_k{ p_k \rho^a_k \otimes \ket{k}\bra{k} },
    \end{equation}
    is the separable state from $A$. More general classical extensions
    (see Definition~\ref{def:ext} in Sect.~\ref{sec:ext_opt}) in which
    $\bar{a}$ is entangled with  $a$ are possible. They would limit,
    though, the {$a$-$b$} correlation-capacity.
        \item The LLA is such that the final state does not exhibit
    any discord with respect to the ancilla:
    ${\delta_{\bar{a}}(\bar{a}:\cdot)=0}$.
        \item Given a classical state, any reduction that preserves
    the bipartition gives rise to a separable state.
    Corollary: it is impossible to find a classical extension of an
    entangled state.
    \end{itemize}
    
The statements above imply that LLA can not be unique, except
for special separable states: those whose convex decomposition of
product states is itself unique, which happens for pure states.
Since separable pure states are product states, they are of no
interest for us here.

A relevant question is whether one can find an {\it optimal} classical
extension of a given separable state, where the vocable `optimal' refers
to some extremal criterion. One could define it, for instance, as being
the classical extension of smallest dimension. We will tackle this
issue with greater precision below and study the relation between
optimality of the classical extension of separable states and their
quantum correlations.

\section{Optimal extension from \textit{separable state} to \textit{classical state}
\label{sec:ext_opt}}
Given a bipartite separable state $\rho^{ab}$ in
${\hilb{ab}=\hilb{a}\otimes\hilb{b}}$, it is always possible to
find a decomposition of the form~\cite{Uhlm98,DVTT00,Lock00}
    \begin{equation} \label{eq:desc_prod}
    \rho^{ab} = \sum_{k=1}^{\lth}{ p_k \ket{a_k}\bra{a_k}
                \otimes \ket{b_k}\bra{b_k} } \,,
    \end{equation}
where ${\rk[\rho^{ab}]\leq \lth\leq(\rk[\rho^{ab}])^2}$. Here, $\lth$
is the states's \textit{cardinality} or \textit{length} and
represents the least number of product states needed for the
purpose. Eq.~\eqref{eq:desc_prod} is the \textit{optimum
decomposition} of $\rho^{ab}$. For separable states in
${\mathbb{C}^2\otimes\mathbb{C}^2}$, one can always find a
decomposition of the type~\eqref{eq:desc_prod} with
${\lth=\max{\{\rk[\rho^{ab}],\rk{[(\rho^{ab})^{Tb}]}\}}\leq4}$,
where $(\rho^{ab})^{Tb}$ is the partial transpose of $\rho^{ab}$~\cite{STV98}.

It is noteworthy that there are other possible decomposition schemes for bipartite states, even in the case of non-separable states. Luo and Sun showed the equivalency of several non-broadcasting theorems using a particular form of bipartite decomposition~\cite{LuSu10}.

Let us introduce some useful definitions:
    \begin{defi} \label{def:ext}
    Given the  bipartite separable state
    $\rho^{ab}$ in ${\hilb{a}\otimes\hilb{b}}$, we say that
    $\sigma^{AB}$ in ${\hilb{A}\otimes\hilb{B}}$ is a
    \textbf{classical extension} of $\rho^{ab}$ if
    \begin{equation} \label{eq:class_ext}
        \trc_{\bar{a},\bar{b}}{[\sigma^{AB}]} = \rho^{ab},
    \end{equation}
and $\sigma^{AB}$ is classical. The partial trace is taken over
$\hilb{\bar{a}}$ and $\hilb{\bar{b}}$, the extensions of
$\rho^{ab}$, with ${\hilb{A}=\hilb{a}\otimes\hilb{\bar{a}}}$ and
${\hilb{B}=\hilb{b}\otimes\hilb{\bar{b}}}$.
    \end{defi}
Here, we could distinguish three possible extensions: from
separable states to CC, CQ or QC states, respectively. As
previously stated, we will be interested in CQ-extensions.
Our following results, though, could be easily generalized
to QC- or CC-extensions.

    \begin{defi} \label{def:ext_opt}
    Given  $\rho^{ab}$ separable in ${\hilb{a}\otimes\hilb{b}}$,
    we say that $\sigma^{AB}$ in ${\hilb{A}\otimes\hilb{B}}$ is
    the \textbf{optimal classical extension} of $\rho^{ab}$ if:
    (a) $\sigma^{AB}$ is a classical extension of $\rho^{ab}$; and
    (b) for any other classical extension $\omega^{A'B'}$ in
    ${\hilb{A'}\otimes\hilb{B'}}$,
    ${\dimn[\hilb{A'}\otimes\hilb{B'}]\geq\dimn[\hilb{A}\otimes\hilb{B}]}$.
    \end{defi}
In general, the \textit{best} Li-Luo's extension is that made from the
optimum decomposition: then the ancilla is ${\mathbb{C}^\lth}$,
with $\lth$ the length of the state to be extended. However, our
Definition~\ref{def:ext_opt} opens the door to possible extensions
not foreseen by the LLA, since it makes no reference
to any particular way of determining the extension. We may have, for instance,
extensions that entangle $a$ with $\bar{a}$. Alternatively, one
may think of extensions that exhibit discord with respect to the
ancilla (i.e., ${\delta_{\bar{a}}(\bar{a}:\cdot)\neq0}$). None of them are
contemplated in the LLA. Consequently, applying LLA to the optimum
decomposition \textit{does not guarantee} an optimal classical extension.
Since we lack a closed formula for the optimum decomposition of arbitrary
separable states, we can not find neither the best Li-Luo's extension
for arbitrary states, nor even less the optimal classical
extension. We show below, however, how to set dimensionality
bounds to our extensions.

\subsection{Bounds for optimal extension}
Theorem~\ref{teo:Luo} says something regarding the dimensionality
of the optimal classical extension. Since ${\rk[\rho^{ab}]\leq
\lth\leq \rk[\rho^{ab}]^2}$, using the optimum decomposition,
Li-Luo's algorithm yields a classical extension for which the
ancilla's dimension is
${d_{\bar{a}}^{Luo}:=\dimn[\mathbb{C}^\lth]=\lth}$, so that
${\rk[\rho^{ab}]\leq d_{\bar{a}}^{Luo}\leq \rk[\rho^{ab}]^2}$. The
optimal classical extension might improve on Li-Luo's, in which case
${d_{\bar{a}}^{opt}<d_{\bar{a}}^{Luo}}$. Regarding our ancilla's
dimension and with regards to bipartite separable states,
the next proposition establishes a general lower bound.
    \begin{prop} \label{prop:ext}
    Let $\rho^{ab}$ be separable in ${\hilb{a}\otimes\hilb{b}}$,
    with length $\lth$, and consider the classical extension $\sigma^{Ab}$
    in ${(\hilb{a}\otimes\hilb{\bar{a}})\otimes\hilb{b}}$, as in
    Definition~\ref{def:ext}. Then, the ancilla's dimension obeys
    \begin{equation} \label{eq:prop_ext}
    d_{\bar{a}} \geq \ceil*{f(d_a,d_b,\lth)} \,
    \end{equation}
where ${{d_x}:=\dimn[\hilb{x}]}$ and ${\ceil*{y}=\min\{n\in\mathbb{Z}|y\leq n\}}$.
The function $f(d_a,d_b,\lth)$ is the only positive root of the quadratic
polynomial ${P_2(x):=c_2 x^2 + c_1 x + c_0}$, with
${c_2:=d_a^2}$, ${c_1:=d_a(d_b^2-1)}$ and ${c_0:=\lth(3-2d_a-2d_b)}$.
    \end{prop}

    \begin{proof}
Let $\lth$ stand for the length of  $\rho^{ab}$ (see
Eq.~\eqref{eq:desc_prod}), this separable state can be expressed
via
    \begin{equation} \label{eq:prop_dem_0}
        \rho^{ab} = \sum_{k=1}^{\lth}{ p_k P^a_k \otimes P^b_k } \,,
    \end{equation}
with ${\{P^a_k\}_{1\leq k\leq \lth}}$ and ${\{P^b_k\}_{1\leq k\leq \lth}}$ projector-sets of rank one
in $\hilb{a}$ and $\hilb{b}$, respectively. The number of independent real
parameters needed for this state's determination is
    \begin{equation} \label{eq:prop_dem_1}
        \lth-1 + \lth(2d_a+2d_b-4) \,.
    \end{equation}
This is obtained as follows. The set  ${\{p_k\}_{1\leq k\leq
\lth}}$ with the condition  ${\sum_k{p_k}=1}$ is determined with
${\lth-1}$ quantities. For each pure state $P^a_k$ one needs
${2d_a-2}$ real parameters. Similar for $P^b_k$. Additionally, given
the classical state $\sigma^{Ab}$ we can cast it as
    \begin{equation} \label{eq:prop_dem_2}
        \sigma^{Ab} = \sum_{m=1}^{d_A}{ q_m \Pi^A_m \otimes \rho^b_m } \,,
    \end{equation}
with ${\{\Pi^A_m\}}$ a basis of rank one orthogonal projectors in $\hilb{A}$,
and ${\{\rho^b_m\}}$ a set of states in $\hilb{b}$.
The index $m$ ranges between $1$ and $d_A=d_ad_{\bar{a}}$.
Accordingly, the set $\{q_m\}$ yields $d_A-1$ independent real
parameters. The set $\{\Pi^A_m\}$ yields ${d_A(2d_A-2)}$
real parameters and we need to discount the ${d_A(d_A-1)}$
restrictions imposed by the commutation rules ${[\Pi^A_m,\Pi^A_n]=0}$,
with ${m>n}$. Note that there are only ${\frac{1}{2}d_A(d_A-1)}$ different
commutation rules, but each complex equation ${[\Pi^A_m,\Pi^A_n]=0}$
counts as two real constraints. In conclusion, ${\{\Pi^A_m\}_{1\leq m\leq d_A}}$ has ${d_A(d_A-1)}$ independent real parameters. Another way to see that ${d_A(d_A-1)}$ is the correct amount of independent real parameters is to take $\{\Pi^A_m\}$ as the rows of a unitary matrix in ${\mathbb{C}^{d_A\times d_A}}$. Such a matrix has $d_A^2$ independent real parameters, but we must subtract $d_A$ arbitrary independent phases, yielding the correct answer.

Also, each $\rho^b_m$ is an arbitrary state of $b$ that is cast as
${\rho_m^b=\sum_s\beta^{(m)}_s \Pi^{b(m)}_s}$ and is determined by
${d_b-1 + d_b(d_b-1)}$ independent real parameters.
Finally, the state $\sigma^{Ab}$ is determined by
    \begin{equation} \label{eq:prop_dem_3}
        d_A^2 + (d_b^2-1)d_A - 1
    \end{equation}
real parameters. The CQ state $\sigma^{Ab}$ requires a
number of parameters greater or equal (Eq.~\eqref{eq:prop_dem_3})
to that for $\rho^{ab}$ (Eq.~\eqref{eq:prop_dem_1}). One ends up
with the above indicated bound for $d_{\bar{a}}$.
    \end{proof}

The following observations are in order.
    \begin{itemize}
        \item The minimum of our bound on $d_{\bar{a}}$~\eqref{eq:prop_ext}
    is always smaller than $\lth$. If ${d_{\bar{a}}^{\min}:=\ceil*{f(d_a,d_b,\lth)}}$ is the minimum
    of~\eqref{eq:prop_ext} for given $d_a$, $d_b$ and $\lth$, and $d_{\bar{a}}^{opt}$ is the unknown theoretical
    minimum for $d_{\bar{a}}$, then
    ${d_{\bar{a}}^{\min}\leq d_{\bar{a}}^{opt}\leq \lth=d_{\bar{a}}^{Luo}}$.
        \item $f(d_a,d_b,\lth)$ grows monotonously with $\lth$, for all
    $\lth\geq1$ and $d_a,d_b\geq1$. Thus, the condition
    ${\rk[\rho^{ab}]\leq \lth\leq \rk[\rho^{ab}]^2}$ establishes both a
    minimum and a maximum to the bound of the proposition,
        \begin{equation} \label{eq:prop_item2}
        \ceil*{f(d_a,d_b,r_{ab})}\leq d_{\bar{a}}^{\min}\leq \ceil*{f(d_a,d_b,r_{ab}^2)} \,,
        \end{equation}
    with $r_{ab}:=\rk[\rho^{ab}]$. For states of maximum rank i.e.,
    $\rk[\rho^{ab}]=d_ad_b$, the bounds depend on the dimensions of
    the parties $a$ and $b$. In particular, in the 2 qubits case one has
    ${\lth=\max{\{\rk[\rho^{ab}],\rk[(\rho^{ab})^{Tb}]\}}}\leq4$.
    Thus, for extending 2 qubits separable states of maximum rank we
    find ${d_{\bar{a}}\geq d_{\bar{a}}^{\min}=2}$.
        \item For full-rank states, $\ceil*{f(d_a,d_b,d_ad_b)}$
    and $\ceil*{f(d_a,d_b,d_a^2d_b^2)}$ are the limit-values for
    $d_{\bar{a}}^{\min}$. Values of $d_{\bar{a}}^{\min}$ are always
    smaller than those obtained via LLA (Tab.~\ref{tab:exts}).
        \begin{table}
        \vspace{.2cm}
        \setlength{\tabcolsep}{8pt}
        \centering
        \begin{tabular}{l c c}
        \hline
        $d$  & ${d_{\bar{a}}^{\min}}$ & ${d_{\bar{a}}^{Luo}}$\\
        \hline\hline
        1       & 1         & 1         \\
        2$^*$   & 2         & 4         \\
        3       & [2,8]     & [9,81]    \\
        4       & [3,13]    & [16,256]  \\
        \hline
        \end{tabular}
        \caption{Ancilla's dimension for the classical extension of
        a bipartite separable state $\rho^{ab}$ with ${d_a=d=d_b}$ and
        maximum rank. 2nd. column: range of values allowed by
        Eq.~\eqref{eq:prop_ext}. 3rd. column: Li-Luo's extension values
        from the optimum decomposition of $\rho^{ab}$. $^*$For $d=2$
        we us the result from Sanpera et al. stating that for these states
        $\lth=4$~\cite{STV98}.}
        \label{tab:exts}
        \end{table}
            \item For systems of greater dimension, consider the case
        ${d_a=d_b=d}$ with full-rank states. From the asymptotic expansion
        of~\eqref{eq:prop_ext} we deduce that
            \begin{equation}
            4 \lesssim d_{\bar{a}}^{\min} \lesssim 2\,d^{3/2} \;(d\rightarrow\infty)\,,
            \end{equation}
        considering that ${r_{ab}\leq\lth\leq r_{ab}^2}$. For these
        states $d^2\leq d_{\bar{a}}^{Luo}=\lth\leq d^4$.
            \item The proposition establishes a lower bound to the
        ancilla's dimensionality in the extension from a separable state
        to a classical-quantum one. If we wished for an classical-classical optimal
        extension, we will deal with a state of the form
        ${\sigma^{AB}=\sum_{m,n}q_{mn}\Pi^A_m\otimes\Pi^B_n}$. The number of
        real parameters of $\sigma^{AB}$ is given by i) ${d_Ad_B-1}$ for
        $\{p_{mn}\}$, ii) ${d_A(d_A-1)}$ for $\{\Pi^A_m\}$,
        and iii) ${d_B(d_B-1)}$ for $\{\Pi^B_n\}$.
        The bounds for $d_{\bar{a}}$ and $d_{\bar{b}}$ are obtained in analogous fashion.
    \end{itemize}
    
From these consideration it follows that, even if the optimal extension remains
unknown, Li-Luo's classical extension, from the optimum decomposition of the
separable state, yields a state that may differ from the one providing the best
classical extension.
We specialize to $2$ qubits next and find more specific results.

\section{Classical extension of separable states in ${\mathbb{C}^2\otimes\mathbb{C}^2}$ \label{sec:2qubits}}
We investigate now possible classical extensions of two-qubits
separable states, with emphasis on states of maximum discord.

\subsection{Extensions in Li-Luo's scheme}
\begin{figure}
    \centering
    \includegraphics[width=8cm]{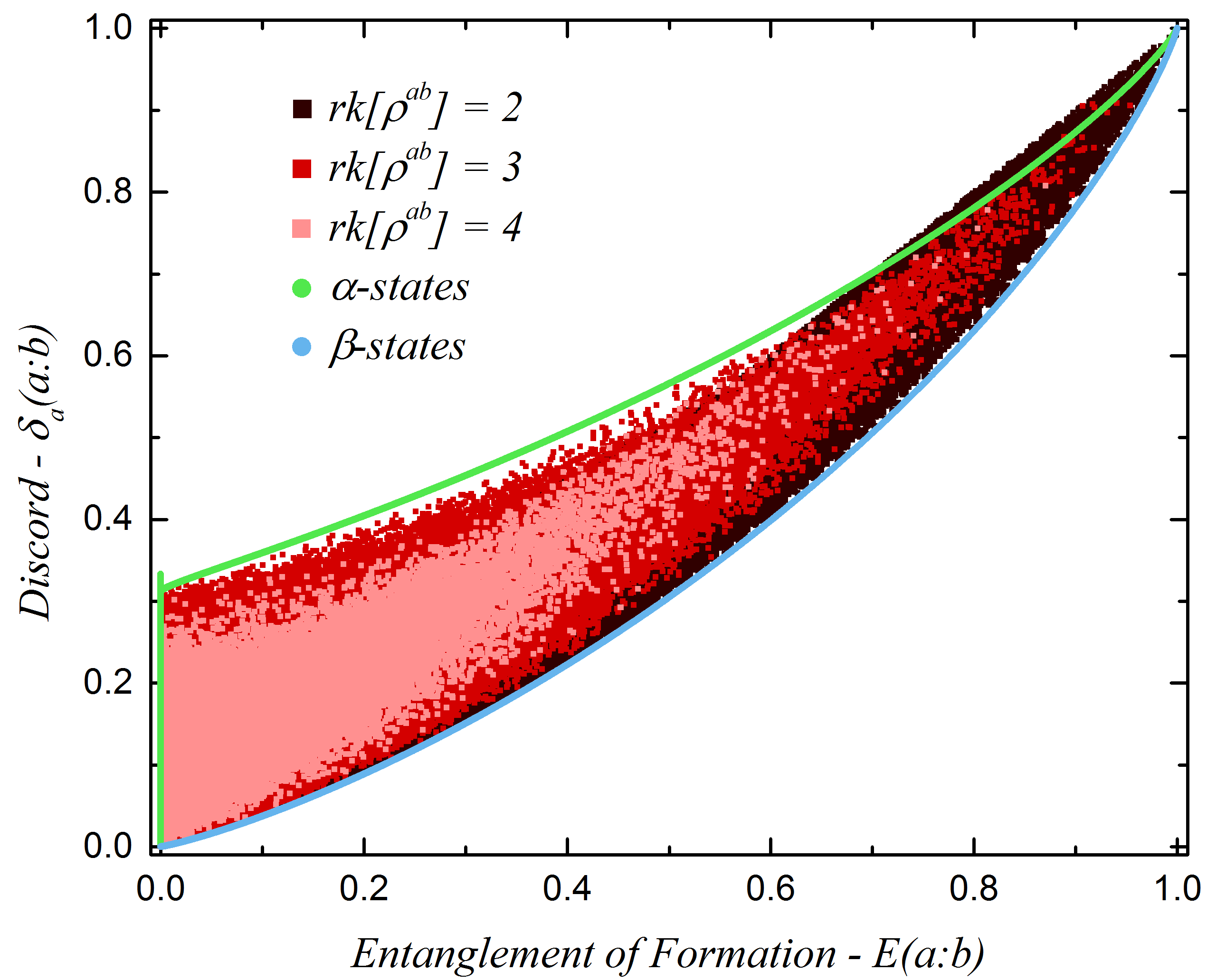}
    \caption{Discord vs. Entanglement of formation for bipartite states
    in ${\mathbb{C}^2\otimes\mathbb{C}^2}$.
    Dots correspond to  ${3\times10^6}$ randomly generated states
    according to Haar's measure. We report results for
      ${1\times10^6}$ rank 2-states, ${1\times10^6}$
      of rank  $3$, and  ${1\times10^6}$ of rank  $4$.
      Green and blue curves correspond, respectively,
      to the families  $\rho_\alpha$ and $\rho_\beta$.}
    \label{fig:discord_flias}
    \end{figure}

\begin{figure}
    \centering
    \includegraphics[width=9cm]{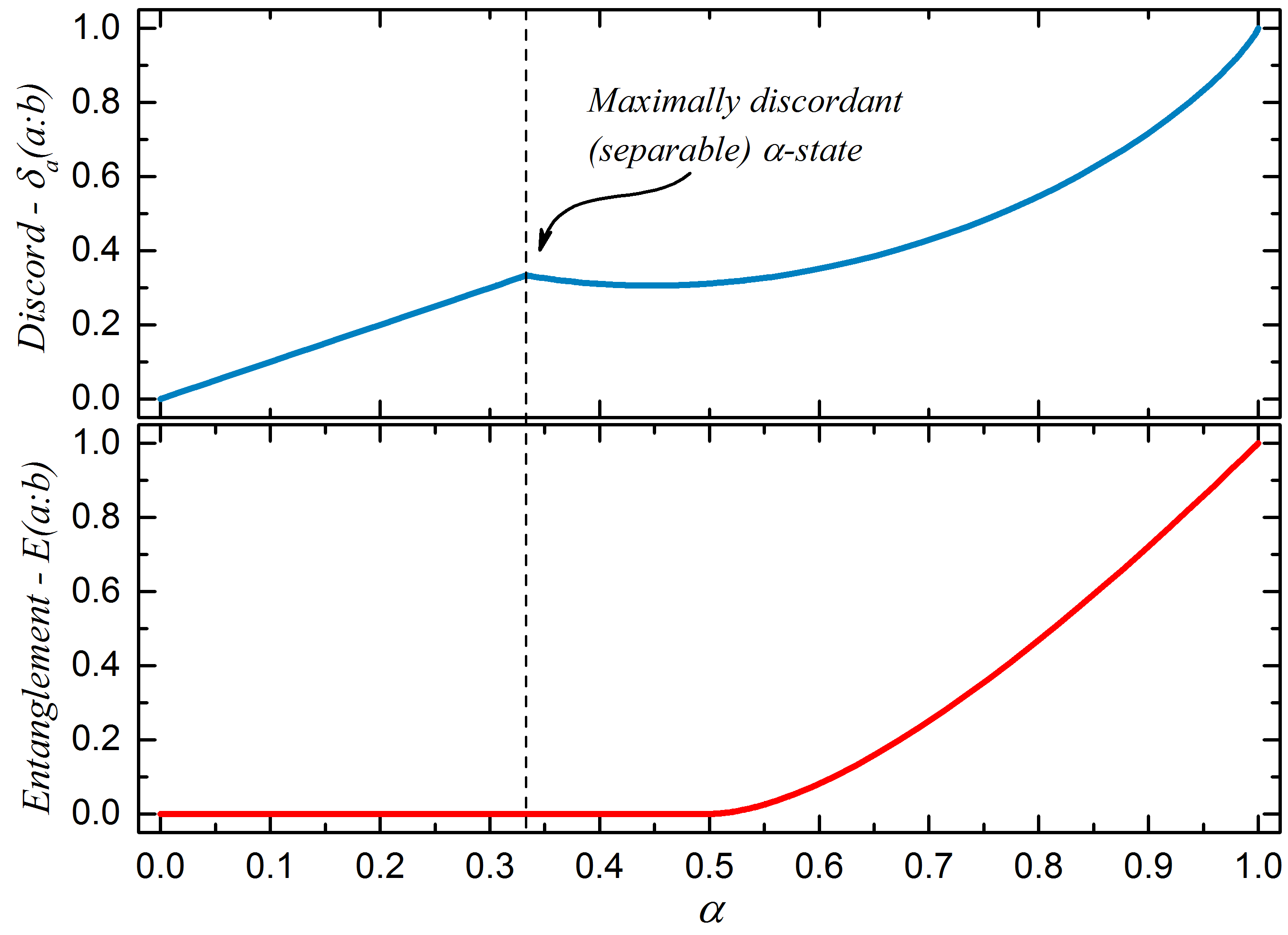}
    \caption{Discord and entanglement of formation for states
        of the family  ${\rho_\alpha}$.}
    \label{fig:alpha_states}
    \end{figure}

In order to find states of maximum discord let us revisit the
relation between discord and entanglement. We are interested in
such states for a fixed rank of the density matrix. In~\cite{Luo08b}, Luo compares the discord and the entanglement of formation for Werner states of two qubits. Moreover, in~\cite{AQJa10}, the authors display such relation for randomly
generated two-qubits states. Fig.~\ref{fig:discord_flias}
reproduces such relation, by numerically computing the discord for
$3\times10^6$ states, and encounter those families that bound
by below and by above the graph discord vs. entanglement.
The family
    \begin{equation} \label{eq:rho_beta}
    \rho(\beta) := \frac{1}{2} \begin{pmatrix}
                    \beta & 0 & 0 & \beta \\
                    0 & 1-\beta & 1-\beta & 0 \\
                    0 & 1-\beta & 1-\beta & 0 \\
                    \beta & 0 & 0 & \beta
                    \end{pmatrix} \,,
    \end{equation}
with  ${0\leq\beta\leq1}$ gives a lower bound for any degree of
entanglement. The states
    \begin{equation} \label{eq:rho_alpha}
    \rho_\alpha := \frac{1}{2} \begin{pmatrix}
                    \alpha & 0 & 0 & \alpha \\
                    0 & 1-\alpha & 0 & 0 \\
                    0 & 0 & 1-\alpha & 0 \\
                    \alpha & 0 & 0 & \alpha
                    \end{pmatrix} \,,
    \end{equation}
with ${0\leq\alpha\leq1}$, give an upper bound for states whose
entanglement ranges between $0$ and $0.620$. For larger
entanglement this limit is provided by Werner states (see
Fig.~\ref{fig:discord_flias})
    \begin{equation} \label{eq:rho_werner}
    \rho_\xi := (1-\xi)\, \frac{\mathds{1}}{4} + \xi \ket{\psi} \bra{\psi} \,,
    \end{equation}
with $-1/3\leq\xi\leq1$ and
$\ket{\psi}:=(\ket{01}-\ket{10})/\sqrt{2}$.
All these families are subsets of the so-called maximally
mixed marginal states, for which an analytical discord-expression is
known. The authors of~\cite{AQJa10} calculate the discord for the
states ${\rho_\alpha}$, finding
    \begin{equation} \label{eq:discord_alpha}
    \delta_a(\alpha) = (1-\alpha)\log(1-\alpha)
        + \alpha\log(\alpha) + (1+\alpha)
        - \frac{1-\bar{\alpha}}{2}\log(1-\bar{\alpha})
        \frac{1+\bar{\alpha}}{2}\log(1+\bar{\alpha}) \,,
    \end{equation}
where ${\bar{\alpha}:=\max{\{|\alpha|,|2\alpha-1|\}}}$. These
states' concurrence is ${C(\alpha)=\max\{0,2\alpha-1\}}$. The
states $\alpha$ are separable for ${\alpha\in(0,\frac{1}{2}]}$.
Of these separable ${\rho_\alpha}$, the one of largest discord
corresponds to ${\alpha=\frac{1}{3}}$. One has
${\delta_a(\rho_\alpha)\rvert_{\alpha=\frac{1}{3}}=\frac{1}{3}}$
(Fig.~\ref{fig:alpha_states}). Note that the optimization can be
achieved in analytic fashion (Cf. Eq.~\eqref{eq:discord_alpha}).
Accordingly, the state
    \begin{equation} \label{eq:max_discord}
    \rho_{\max}^{\lth=4} := \rho_\alpha\bigr\rvert_{\alpha=\frac{1}{3}}
        = \frac{1}{6} \begin{pmatrix}
                    1 & 0 & 0 & 1 \\
                    0 & 2 & 0 & 0 \\
                    0 & 0 & 2 & 0 \\
                    1 & 0 & 0 & 1
                    \end{pmatrix}
    \end{equation}
is representative of maximum discord-separable states in
${\mathbb{C}^2\otimes\mathbb{C}^2}$. We have
${\rk{[\rho_{\max}]}=3}$ and ${\rk{[\rho_{\max}^{Tb}]}=4}$, so one
expects to find a separable decomposition of the type~\eqref{eq:desc_prod},
with $\lth=4$. Thus, ${\rho_{\max}^{\lth=4}}$ can be classically extended
via LLA with ${d_{\bar{a}}^{Luo}=4}$. On the other hand, it is possible to
find classical states of smaller dimension whose separable reductions
reaches a discord-amount close to the maximum. For instance, the state
    \begin{equation} \label{eq:max_discord_2}
    \rho_{\max}^{\lth=3} :=
    \rho_\alpha\bigr\rvert_{\alpha=\frac{1}{2}}
        = \frac{1}{4} \begin{pmatrix}
                    1 & 0 & 0 & 1 \\
                    0 & 1 & 0 & 0 \\
                    0 & 0 & 1 & 0 \\
                    1 & 0 & 0 & 1
                    \end{pmatrix}
    \end{equation}
has a discord that equals $93\,\%$ of the discord accrued to the
state ${\rho_{\max}^{\lth=4}}$ and can be classically extended
with ${d_{\bar{a}}^{Luo}=3}$. Similarly, the state
    \begin{equation} \label{eq:max_discord_3}
    \rho_{\max}^{\lth=2} := \frac{1}{2} (\ket{0}\bra{0}\otimes\ket{0}\bra{0}
                            + \ket{+}\bra{+}\otimes\ket{1}\bra{1}) \,,
    \end{equation}
that can be classically extended with  ${d_{\bar{a}}^{Luo}=2}$,
exhibit a discord equal to  $61\,\%$ of that of
${\rho_{\max}^{\lth=4}}$ (see Tab.~\ref{tab:discords}).
\begin{table}
    \vspace{.2cm}
    \setlength{\tabcolsep}{7pt}
    \centering
    \begin{tabular}{c c}
    \hline
    ${\lth\,(=d_{\bar{a}}^{Luo})}$  & $\delta_a(a:b)$ \\
    \hline\hline
    4           & $\frac{1}{3} \approx 0.3333$    \\
    3           & $(\frac{3}{4})\log{(\frac{4}{3})} \approx 0.3113$  \\
    2           & $2-(\frac{\surd 2}{2})\log(3+2\surd 2) \approx 0.2018$ \\
    1           & $0$\\
    \hline
    \end{tabular}
    \caption{Discord for maximally discording separable states in
     ${\mathbb{C}^2\otimes\mathbb{C}^2}$ according to their length.}
    \label{tab:discords}
\end{table}

\vspace{.4cm}\noindent
\textit{Separable decomposition of ${\rho_{\max}^{\lth=3}}$.}\\
We continue with the issue of expressing, for different ranks,
states of large discord. For ${\lth=3}$ the maximum discord is
$0.3113$, a value reached by the state ${\rho_{\max}^{\lth=3}}$
of Eq.~\eqref{eq:max_discord_2}. It's easy to verify that
    \begin{equation} \label{eq:rk3_maxdis}
    \rho_{\max}^{\lth=3} \eqU \frac{1}{4} \left( P_0{\otimes}P_0 + P_1{\otimes}P_1
                        + P_+{\otimes}P_+ + P_-{\otimes}P_- \right) \,
    \end{equation}
with ${\{P_j\}_{j=0,1,+,-,r,l}}$ the eigen-projectors of
$\sigma_z$ and $\sigma_x$, respectively. By `$\eqU$' we indicate an
equivalence up to unitary transformations. Eq.~\eqref{eq:rk3_maxdis}
is a possible separable decomposition, but it is not optimal.
To find the separable optimum decomposition of a given bipartite state one
proceeds as described in~\cite{STV98}. Denoting by
    \begin{equation} \label{eq:desc_sep_1}
    \ket{\theta,\phi} := \cos\left(\frac{\theta}{2}\right) \ket{0}
                        + \exp(i \phi)\sin\left(\frac{\theta}{2}\right) \ket{1} \,,
    \end{equation}
an arbitrary pure state in $\mathbb{C}^2$, we find that the set
${\mathcal{W}=\{\ket{0,0},\ket{\frac{2\pi}{3},0},\ket{\frac{2\pi}{3},\pi}\}}$
defines the optimum decomposition
    \begin{equation} \label{eq:rk3_max_opt}
    \rho_{\max}^{\lth=3} \eqU \frac{1}{3}\sum_{i=1}^3{W_k\otimes W_k} \,
    \end{equation}
with ${W_k=\ket{w_k}\bra{w_k}}$ and ${\ket{w_k}\in\mathcal{W}}$.
We repeat things below for ${\rho_{\max}^{\lth=4}}$.

\vspace{.6cm}\noindent
\textit{Separable decomposition of ${\rho_{\max}^{\lth=4}}$.}\\
It is easy to see that ${\rho_{\max}^{\lth=4}\eqU\rho_\alpha}$,
with ${\alpha=\frac{1}{3}}$, and that it can be decomposed as
    \begin{equation} \label{eq:rk4_dis_max}
    \rho_{\max}^{\lth=4} \eqU \frac{1}{6}
    \left( P_0{\otimes}P_0 + P_1{\otimes}P_1 + P_+{\otimes}P_+
        + P_-{\otimes}P_- + P_r{\otimes}P_r + P_l{\otimes}P_l \right) \,,
    \end{equation}
with ${\{P_j\}_{j=0,1,+,-,r,l}}$ the eigen-projectors of
$\sigma_z$, $\sigma_x$ and $\sigma_y$, respectively. We seek now
for the optimum decomposition. For simplicity's sake, instead of
${\rho_{\max}^{\lth=4}}$ we consider
    \begin{equation} \label{eq:rho_max_2}
    \tilde{\rho}_{\max} := \frac{1}{6} \begin{pmatrix}
                    2 & 0 & 0 & 0 \\
                    0 & 1 & 1 & 0 \\
                    0 & 1 & 1 & 0 \\
                    0 & 0 & 0 & 2
                    \end{pmatrix} \,,
    \end{equation}
obtained from ${\rho_{\max}^{\lth=4}}$ via a local (in $b$)
unitary transformation, which does not change the discord. This
transformation consists of a  swap in $b$,
    \begin{equation} \label{eq:swap_b}
    U_b := \begin{pmatrix}
                    0 & 1 \\
                    1 & 0
                    \end{pmatrix} \,,
    \end{equation}
such that ${\tilde{\rho}_{\max}=U\rho_{\max}^{\lth=4}U^\dagger}$,
with ${U:=\mathds{1}_a\otimes U_b}$ and $\mathds{1}_a$ the identity
in $a$.

Defining ${\mathcal{Z}=\{\ket{0,0},\ket{\theta^*,0},
\ket{\theta^*,\frac{2\pi}{3}},\ket{\theta^*,\frac{4\pi}{3}}\}}$,
with ${\theta^*=\arccos(-\frac{1}{3})}$, the optimum decomposition
of ${\tilde{\rho}_{\max}}$ is
    \begin{equation} \label{eq:desc_sep_2}
    \tilde{\rho}_{\max} = \frac{1}{4} \sum_{k=1}^4{Z_k\otimes Z_k} \,,
    \end{equation}
with ${Z_k=\ket{z_k}\bra{z_k}}$ and ${\ket{z_k}\in\mathcal{Z}}$.

Note that all states exhibit the same overlap among themselves,
i.e., that ${|\braket{\theta_k,\phi_k|\theta_{k'},\phi_{k'}}|^2=c}$,
${\forall k\neq k'}$ (${c=1/3}$). In terms of a parameterization of
states on the  Bloch sphere, where
${r_k=(\sin(\theta_k)\cos(\phi_k),\sin(\theta_k)\sin(\phi_k),\cos(\theta_k))}$
is the position-vector associated to ${\ket{\theta_k,\phi_k}}$,
the angle between two different states is always $2\pi/3$.
Summing up, the pure states in $\tilde{\rho}_{\max}$
can be associated to the four vertices of the regular 3-simplex in a
three-dimensional space (a tetrahedron).
As shown by Eq.~\eqref{eq:desc_sep_2}, both subsystems have the same
pure states. Thus, the state of maximum discord for 2
qubits can be expressed as the equal-weights, convex combination of
$4$ symmetric product states given by $4$ pure states that are
maximally distinguishable.
Given this states' symmetry in both qubits' spaces, any choice of
projective measurement will yield the same discord.

\subsection{Optimal classical extensions}
\begin{figure}
    \centering
    \includegraphics[width=7cm]{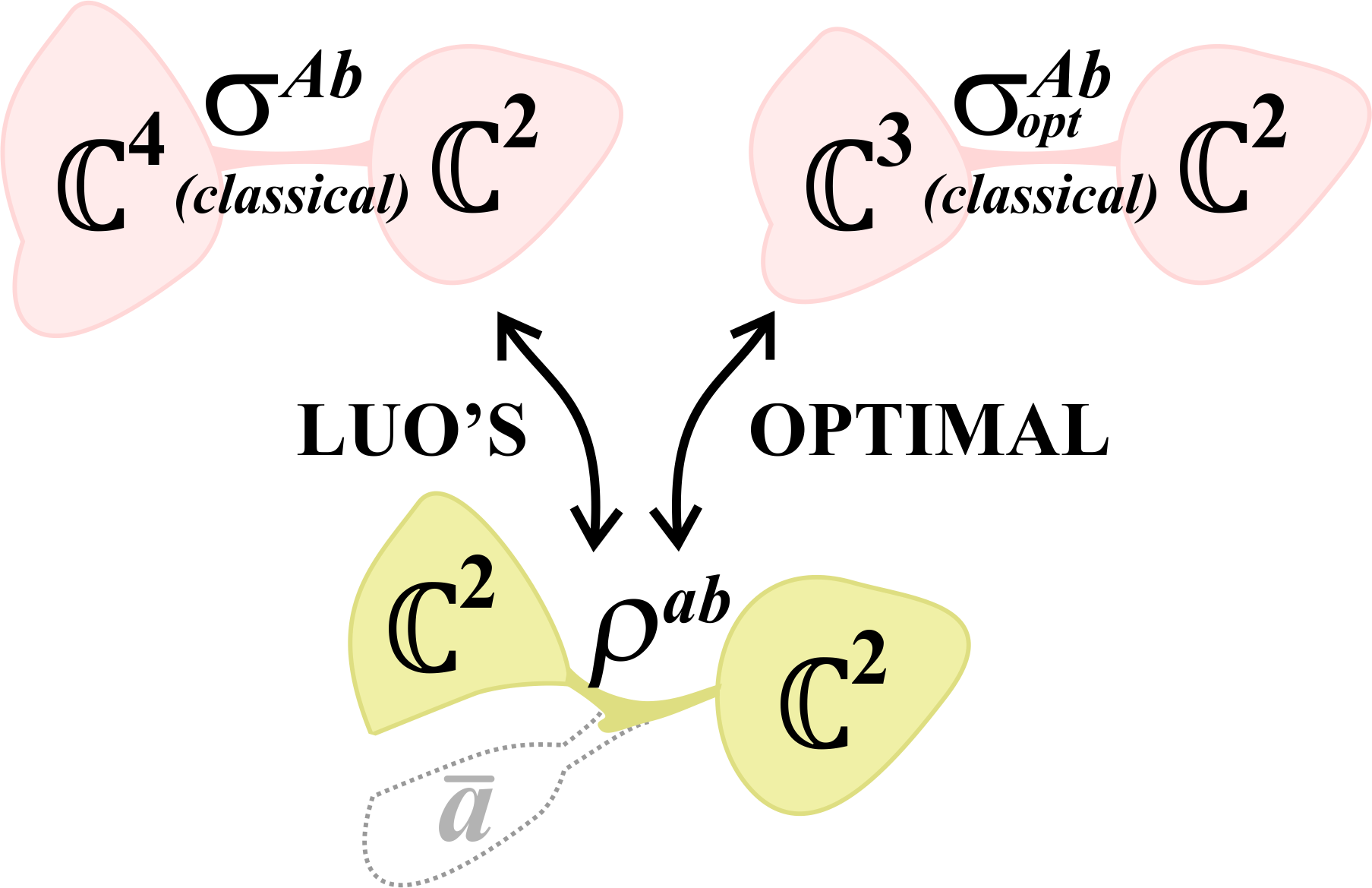}
    \caption{So as to extend the 2 qubits, maximally discordant separable state
    ${\rho_{\max}^{\lth=4}}$,
    Li-Luo's algorithm employs an ancilla in $\mathbb{C}^4$.
    In the optimal scheme, it is possible
    to find a compatible extension in $\mathbb{C}^3$.}
    \label{fig:ext_opt}
    \end{figure}

As suggested by table~\ref{tab:exts}, it is possible to improve on
the results of the LLA. Notice from
Fig.~\ref{fig:ext_opt} that it is possible to classically extend
${\rho_{\max}^{\lth=4}}$ with a qutrit, while the LLA needs
2 qubits. Similarly, we can extend
${\rho_{\max}^{\lth=3}}$ with 1 qubit, versus the 1 qutrit
required by the LLA.
These new extensions were numerically obtained via Monte Carlo so
as to find the reductions in ${\mathbb{C}^2\otimes\mathbb{C}^2}$
[of classical states in ${\mathbb{C}^{d_A}\otimes\mathbb{C}^2}$]
of largest discord. One starts building up a classical state
${\sigma_{Ab}=\sum_k{p_k\Pi_k^A\otimes\rho_k^b}}$, with
${\{\Pi_k^A\}_{1\leq k\leq d_A}}$ orthonormal projectors of
$\mathbb{C}^{d_A}$. The family of orthonormal projectors is
obtained as the columns of an arbitrary  unitary matrix
${U_{A}\in\mathbb{C}^{d_A\times d_A}}$. The 4 states ${\rho_k^b}$
are arbitrary in $\mathbb{C}^2$ and $\{p_k\}$
a probability distribution. Given the prevailing symmetry in the
maximally discordant states, we choose
${\rho_k^b=\rho_k^a=\trc_{\bar{a}}\Pi_k^A}$ and ${p_k=\frac{1}{d_A}}$
for all $k$, so that the classical state becomes determined solely by $U_A$.
This is the only element that varies in each algorithm's step,
which considerably simplifies computations.
    \begin{figure*}
    \centering
    \includegraphics[width=15.5cm]{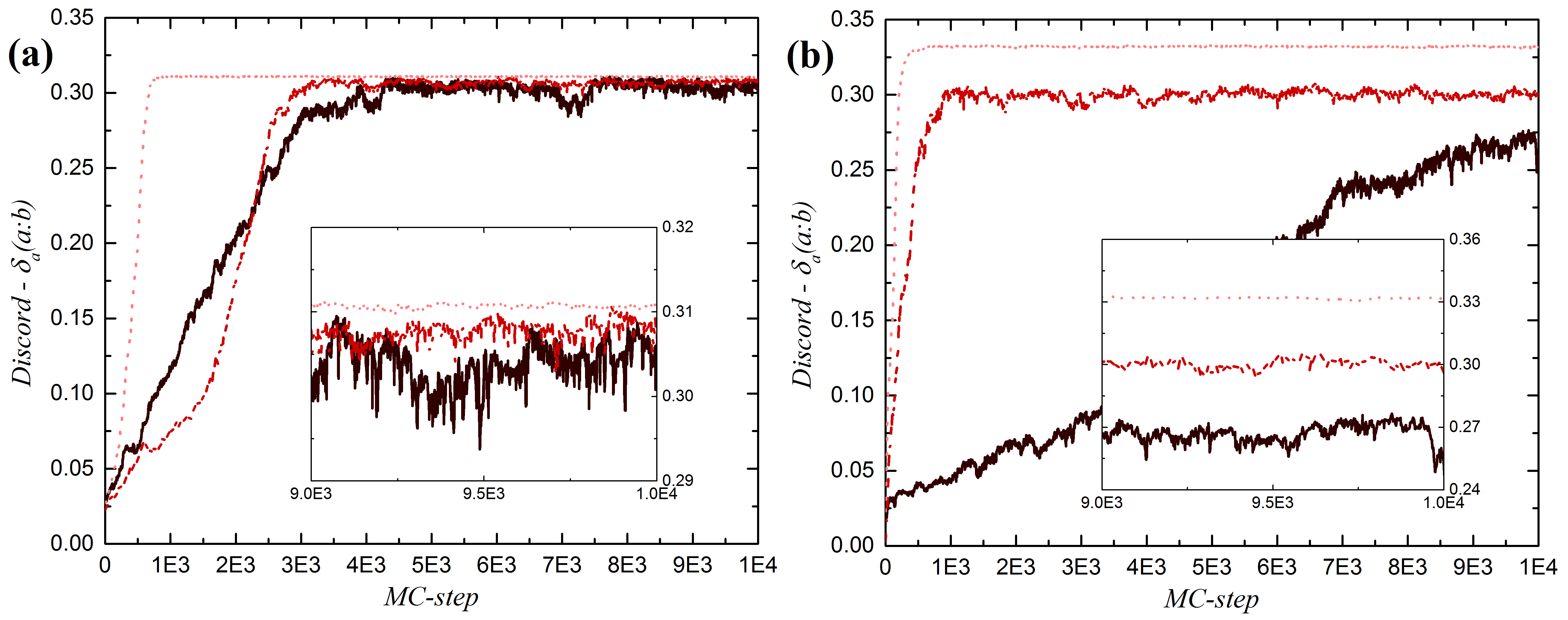}
    \caption{Search for the maximally discordant separable states of
    2 qubits, obtained via reductions of classically correlated
    states using the Monte Carlo method.
    Each line corresponds to a different simulation-temperature.  (a)
    Using classical states in  ${\mathbb{C}^4\otimes\mathbb{C}^2}$
    one finds reductions whose maximum discord is ${\delta_a(a:b)=0.3113}$.
    (b) Using classical states in ${\mathbb{C}^6\otimes\mathbb{C}^2}$
    one finds reductions with maximum discord ${\delta_a(a:b)=0.3333}$.}
    \label{fig:ext_mc}
    \end{figure*}

Fig.~\ref{fig:ext_mc} displays our results. The maximally discordant
separable state, with ${\delta_a(a:b)=0.3333}$, is obtained
as the reduction of a classical state  with $d_A=3$. For $d_A=2$,
the reductions' maximum discord is seen to be
${\delta_a(a:b)=0.3113}$. The columns of the unitary matrix
    \begin{equation} \label{eq:UA_max}
    \text{\scriptsize%
    $U_A^{opt} = \begin{pmatrix}%
    0.5288-0.2428\,i & -0.0241+0.0541\,i & 0.2730-0.0396\,i & 0.5695+0.3689\,i & -0.1512-0.1230\,i & 0.2672-0.1097\,i \\%
    -0.0179+0.2237\,i & 0.1392+0.1287\,i & 0.1575-0.8817\,i & -0.2307+0.1243\,i & -0.1110-0.0388\,i & 0.1259-0.1150\,i \\%
    -0.0670+0.1750\,i & -0.0525-0.0246\,i & 0.0387+0.2783\,i & -0.2118+0.0457\,i & -0.4907+0.1406\,i & 0.1647-0.7403\,i \\%
    0.4663+0.4930\,i & 0.0701+0.2679\,i & -0.0392+0.0417\,i & 0.1412-0.5644\,i & 0.2635+0.1158\,i & 0.1552-0.1193\,i \\%
    -0.2532+0.0657\,i & 0.8938+0.0569\,i & 0.0919+0.1655\,i & 0.1642+0.1537\,i & 0.1726+0.0124\,i & 0.0627-0.0954\,i \\%
    -0.2169-0.0076\,i & -0.2708+0.0706\,i & -0.0449-0.0414\,i & 0.0949+0.1610\,i & 0.6485-0.3928\,i & -0.0244-0.5103\,i%
    \end{pmatrix}$ }
    \end{equation}
determine, on the standard basis, the basis ${\{\Pi_k^A\}_{1\leq
k\leq 6}}$ of the classical state $\sigma^{Ab}_{opt}$ such that
$\trc_{\bar{a}}\sigma^{Ab}_{opt}$ exhibits maximum discord:
${\delta_a(a:b)=0.3333}$. We are finding an extension in
${\mathbb{C}^6\otimes\mathbb{C}^2}$ of ${\rho_{\max}^{\lth=4}}$,
improving on the LLA. Why is this extension unattainable in
${\mathbb{C}^6\otimes\mathbb{C}^2}$ via the Li-Luo's approach? It
suffices to note that  ${\lth=4}$, so that the LLA demands an ancilla
in $\mathbb{C}^4$ so as to classically
extend things to ${\rho_{\max}^{\lth=4}}$. We conjecture that
$\sigma^{Ab}_{opt}$ is the optimal extension of
${\rho_{\max}^{\lth=4}}$.

Notice the following difference between Li-Luo's extension and the
optimal one. In the later, the ancilla is correlated only with
the set $ab$, but not individually with $a$ or $b$, i.e., 
${I(\bar{a}:a)=0}$ and ${I(\bar{a}:b)=0}$ but ${I(\bar{a}:ab)=0.585}$.
Instead, for Li-Luo's extension, one has ${I(\bar{a}:a)=1}$,
${I(\bar{a}:b)=1}$, and ${I(\bar{a}:ab)=1.585}$.

\section{Maximally discordant separable states \label{sec:mdss}}
The previous results in ${\mathbb{C}^2\otimes\mathbb{C}^2}$
suggest that maximally discordant separable states (MDSS) posses a
rank close to the maximum. We see next how some symmetries
associated to the construction of maximally discordant separable
states of 2 qubits can be generalized to spaces of greater dimension.

Eqs.~\eqref{eq:rk3_maxdis} and~\eqref{eq:rk4_dis_max} indicate
that 2 qubits MDSS can be built by uniformly mixing
states corresponding to different \textit{mutually unbiased bases}
(MUBs). Indeed, $\rho_{\max}^{\lth=3}$ is constructed mixing two
MUBs ($\sigma_z$ and $\sigma_x$ in our example) and
$\rho_{\max}^{\lth=4}$ is erected mixing the 3 possible MUBs.
We look now for a possible generalization of these MDSS to
arbitrary dimension.

For {${d\times d}$-dimensional} states, if
${\{P_k^i\}_{1\leq k\leq d+1}^{1\leq i\leq d}}$ is the set of
projectors determining the ${d+1}$ MUBs of one of the parties,
the state
    \begin{equation} \label{eq:mdss1}
    \rho_{\max}^d := \frac{1}{d(d+1)}\sum_{k=1}^{d+1}\sum_{i=1}^{d}{P_k^i\otimes P_k^i} \,.
    \end{equation}
should be a plausible candidate of a maximally discordant state.

Another possible MDSS-generalization (Eq.~\eqref{eq:rho_max_2})
to larger dimensions starts from noting that the projectors basis
of rank $1$ ${\{Z_k\}_{1\leq
k\leq4}}$ of Eq.~\eqref{eq:desc_sep_2} constitutes a
\textit{symmetric and informationally complete positive operator
valued measure} (SIC-POVM) in $\mathbb{C}^2$. In fact, taking
$E_k:=Z_k/d$ and $d=2$ one has
    \begin{equation} \label{eq:sic_povm}
    \sum_{k=1}^{d^2}{E_k} = \mathds{1} \,,
    \end{equation}
and
    \begin{equation} \label{eq:sic_povm2}
    \trc(E_k E_{k'}) = \frac{1}{d^2(d+1)} \;, k\neq k' \,.
    \end{equation}
Equivalently,
    \begin{equation} \label{eq:sic_povmb}
    \frac{1}{d}\sum_{k=1}^{d^2}{Z_k} = \mathds{1} \,,
    \end{equation}
and
    \begin{equation} \label{eq:sic_povm_2b}
    \trc(Z_k Z_{k'}) = \frac{1}{d+1} \;, k\neq k' \,.
    \end{equation}
    In the $d$-dimensional case, a SIC-POVM is a
set ${\{Z_k\}_{1\leq k\leq d^2}}$ of rank $1$ projectors
obeying~\eqref{eq:sic_povmb}--\eqref{eq:sic_povm_2b}.
A trivial generalization to two qudits is given by the state
    \begin{equation} \label{eq:mdss2}
    \tilde\rho_{\max}^d := \frac{1}{d^2}\sum_{k=1}^{d^2}{Z_k\otimes Z_k} \,.
    \end{equation}

The existence of  SIC-POVMs in $\mathbb{C}^{d}$ has not been
demonstrated yet for arbitrary  $d$, although it is proved for
$d$ prime or $d$  a power of a prime. Our problem is
equivalent to that of finding $d^2$ rays separated by equal angles in
${\mathbb{C}^{d}}$~\cite{Kibl13,Kibl14}, being intimately linked
to the  existence of  $d+1$ mutually unbiased bases (MUBs) in
$\mathbb{C}^d$ and thus with the existence of complementary
observables~\cite{ApDF07,Stac14}. Alternatively, our problem can
be seen as that of embedding the simplex {($d^2$-$1$)-dimensional}
generated by $d^2$ pure states into the  convex of quantum states
in such a way that all pure states exhibit the same
\textit{overlap}~\cite{MMPZ08}. This is the way in which we
interpret the tetrahedron formed by the components of
${\tilde{\rho}_{\max}}$ in Eq.~\eqref{eq:desc_sep_2}. A SIC-POVM
is that  POVM that better approximates an orthonormal basis in the
states-space~\cite{Scot06}. It is interesting to note that recently
some authors introduced a new measure of quantum correlations
involved in the optimal acquisition of information
over all the local MUBs~\cite{WMCY14}.

\subsection{Genuine quantum correlations \label{sec:gen}}
Recent works show that one can obtain states with finite discord by
effecting local operations on states of null discord~\cite{Gess12,Gior13}.
Thus, one may view discord as a resource,
\textit{necessary}, but not sufficient, to attain genuine quantum
correlations. A way of point out toward states with genuine
quantum correlations is through their decomposition in
product states of local bases~\cite{DaVB10}. If $\{A_m\}$ and
$\{B_n\}$ are bases associated to Hermitic operators in $\hilb{A}$
and $\hilb{B}$, respectively, the composite states $\sigma^{AB}$
can be decomposed as
    \begin{equation} \label{eq:local_desc}
    \sigma^{AB} := \sum_{m=1}^{d_A^2}\sum_{n=1}^{d_B^2}{r_{mn} A_m\otimes B_n } \,,
    \end{equation}
with  $d_A$ ($d_B$) the dimension of $\hilb{A}$ ($\hilb{B}$). The
\textit{correlation matrix} $R:=(r_{mn})$ can be recast via
decomposition in singular values.  If ${L_R:=\rk[R]}$ is its rank
and  $s_l$ its singular values,
    \begin{equation} \label{eq:svd_desc}
    \sigma^{AB} := \sum_{l=1}^{L_R}s_l F_l\otimes G_l \,,
    \end{equation}
where  $F_l$ and $G_l$ are the  elements of $A$ and $B$,
respectively, in the new basis. If the states-components are pure,
${L_R\leq(\dimn[\hilb{ab}])^2}$ (Cf. Eq.~\eqref{eq:desc_prod}).
If not (mixed states allowed) one has
${L_R\leq d_{\min}^2}$, where ${d_{\min}:=\min\{\dimn[\hilb{a}],\dimn[\hilb{b}]\}}$
corresponds to that subsystem of smaller dimension. For classical states $L_R$
is bound (by above) by the dimension of the subsystems, i.e., ${L_R\leq d_{\min}}$.
There are states of finite discord with ${L_R\leq d_{\min}}$, but one can show that their
discord can be created vial local operations, so that they do not constitute quantum
resources~\cite{Gess12,Gior13}. States with ${L_R>d_{\min}}$ have discord
necessarily and their correlation matrix is not compatible with that pertaining
to a classical state. Only these states are genuinely quantum (with respect
to their correlations). Summing up, $L_R$ is the signature of quantum-correlated
states that can not be obtained from classical states via local operations.

For instance, if ${\tilde{\rho}_{\max}}$, Eq.~\eqref{eq:desc_sep_2}
represents the decomposition~\eqref{eq:svd_desc}, with
${\{\ket{\theta_l,\phi_l}\bra{\theta_l,\phi_l}\}_{1\leq l\leq4}}$
the basis of hermitic operators both in $\hilb{a}$ and $\hilb{b}$,
and  ${s_l=\frac{1}{4}\;\forall\,l}$. Here, the correlation matrix
is of rank $4$. Also, ${d_{\min}=\dimn[\mathbb{C}^2]=2}$. Thus,
${L_R>d_{\min}}$ and the correlations are indeed genuinely
quantum. On the other hand, it is easy to see that for the state
in Eq.~\ref{eq:max_discord_3} the correlation matrix
is of rank $2$. Discord-like correlations can here be locally
created. As a corollary, for 2 qubits bipartite states, genuinely
quantum states with discord are only those of ${L_R>2}$. In
Tab.~\ref{tab:discords}, only the states with ${\lth>2}$ are
relevant.

Note that, given our decomposition~\eqref{eq:mdss2} of
${\rho_{\max}^d}$, since the $\{M_k\}$ are linearly independent,
the number of terms automatically determines  the rank of the
correlation matrix. Here one has ${\rk[R]=d^2>d_{\min}}$~\cite{Chen13},
since ${d_{\min}=\dimn[\mathbb{C}^d]=d}$. Thus,
for these states the discord is not spurious in the sense
discussed above. In other words, for any dimension, states that
are separable and possess discord defined via Eq.~\eqref{eq:mdss2}
constitute genuine quantum resources.

\section{Conclusions \label{sec:conc}}
Summarizing our results:
    \begin{itemize}
        \item We have demonstrated in this work that the existence of
genuine quantum correlations in separable states is related to the
possibility of extending such states to classically correlated
ones of larger dimension.
        \item We have introduced the notion of optimum classical extension of
separable states and showed that the algorithm advanced by Li and
Luo can be, in general, improved.
        \item We also found that the maximum degree of discord of a
given separable state is linked to the dimensionality of its optimum
classical extension.
        \item We demonstrated the existence of a lower bound for the
dimension of such extension.
        \item For 2 qubits separable states we found different classical
extensions for states of maximum discord. In particular, we showed that
with one qutrit we can classically extend the 2 qubits state of maximum
discord. On the basis of numerical simulations we conjectured that such
a classical extension is the optimum one.
        \item Our results for low dimensionality systems induce hypothesis
concerning the structure of separable states of maximum discord in
arbitrary dimension that, in turn, suggest interesting links involving the
notions of mutually unbiased basis and symmetric and informationally complete
positive operator valued measures (SIC-POVMs).
    \end{itemize}

\bibliography{ref}

\end{document}